\documentstyle[aps,preprint]{revtex}
\begin{document}
\draft
\title{Charge-monopole versus Gravitational Scattering at Planckian Energies}
\author{Saurya Das and Parthasarathi Majumdar}
\address{The Institute of Mathematical Sciences, CIT Campus, Madras
600113,  India.}
\maketitle
\begin{abstract}
The amplitude for the scattering of a point magnetic monopole and a
point charge, at centre-of-mass energies much larger than the masses of
the particles, and in the limit of low momentum transfer, is shown to
be proportional to the (integer-valued) monopole strength,
assuming the Dirac quantization condition for the monopole-charge system.
It is demonstrated that, for small momentum transfer, charge-monopole
electromagnetic effects remain comparable to those due to the gravitational
interaction between the particles even at Planckian centre-of-mass energies.
\end{abstract}
\pacs{14.80.Hv, {\bf 12.25.+e}}

The amplitude due to graviton exchange between two particles colliding
at Planckian centre-of-mass (cm) energies and in the limit of
vanishingly small momentum transfer has been shown \cite{thf},\cite{amv} to
be calculable semi-classically, corresponding to the eikonal
approximation of quantum gravity. This simplification originates from
the decoupling of all physical graviton modes in this kinematical
regime, and the pure gauge states left behind have been shown to be
describable by a topological sigma model in 1+1 dimensions with
diffeomorphism invariance \cite{ver} resembling a string theory. The
semi-classical amplitude found by 't Hooft also showed a striking
resemblance to the Veneziano amplitudes of string theory. One
significant aspect was the existence of poles in this amplitude at cm
energies $Gs = -iN$ (where $N$ is a positive integer) \cite{thf2}.
If the particles carry electric charge, the
amplitude is still calculable semi-classically, corresponding to a
shift $Gs \rightarrow Gs + \kappa \alpha$ with $G$ being Newton's
constant, $s$ the squared cm energy, $\kappa$ is any integer (positive or
negative) and $\alpha$ is the fine structure constant. Thus, at the
Planck scale, electromagnetic effects due to these charges constitute
a minor perturbation on the gravitational effects. At sub-Planckian cm
energies that are still
large compared to the particle masses, the quantum electrodynamic eikonal
approximation \cite{suc} has been exactly reproduced from the semi-classical
amplitude, and the corresponding topological sigma model
has been identified \cite{jac}.

In this paper, we study the scattering of a Dirac magnetic monopole
with a pointlike electric charge in the same kinematical limit. At
sub-Planckian
energies (i.e., in the absence of graviton exchange), the amplitude
turns out, remarkably, to be proportional to the integer $n$ occuring
in the Dirac quantization condition, modulo standard kinematical
factors. Next we turn on gravity, and show that even when the cm energy
becomes on the order of the Planck scale, the monopole-charge
interaction contribution to the cross section remains almost as
large as the gravitational contribution, the former being
characterised by the monopole strength $n$ as compared to the energy
dependent gravitational coupling strength $Gs$ occuring in the latter.
Another remarkable feature of the amplitude in the presence of
magnetic charge is the existence of poles at $Gs = -i(N + {n \over 2})$,
in contrast to poles at $Gs = -iN$ for two particle scattering, and
akin to poles in {\it three} particle scattering amplitudes
found by 't Hooft \cite{thf2}. We
shall comment on this result towards the end of the paper.

Consider first the scattering of an ultra-relativistic monopole from a
slowly-moving charged particle within a kinematical region defined by
a cm energy $s~>>~m^2_{1,2}$ ($m_{1,2}$ are the rest masses of the
particles) and momentum transfer $t~\rightarrow~0$, where, $s$ and $t$
are the usual Mandelstam variables. This kinematical region
corresponds to a situation where the particles scatter almost solely
in the forward direction, and for all practical purposes behave as
though they are massless \cite{thf}. Further, the scattering can be
described in terms of the response of the `target' charged particle to
the electromagnetic `shock' wave-front carried by the fast-moving monopole.
Treating the shock wave-front as classical, the calculation of the
scattering amplitude reduces to determining the overlap of the
wavefunctions of the point charge evaluated before and after
encountering the monopole shock front. Basically, the fields in the
shock front produce a phase factor in the charge wave-function,
which can be exactly computed.

To determine the fields due to an ultra-relativistic monopole, we first
boost the vector potential of a static monopole along the positive
$z$-axis to some velocity $\beta$, and then evaluate the limit of this
potential as $\beta \rightarrow 1$, as in ref.s {\cite{thf}} and
{\cite{jac}}. The vector potential of a static monopole carrying
magnetic charge $g$ can be given in spherical polar coordinates,
following Wu and Yang \cite{wuy}, by
\begin{eqnarray}
{\vec {A}}~^I ~& = &~ {g \over {r sin \theta} } (1 - cos
\theta) {\hat \phi}~,~~~~~  0 \le \theta < \pi \label{vec1}  \\
{\vec{A}}~^{II}~& = &~{-~g \over { r sin \theta}}
(1+cos\theta){\hat \phi}~,~~~~~  0 < \theta \le \pi~.\label{vec2}
\end{eqnarray}
Observe that we have made the gauge choice $A^0=0$, and have chosen an
orientation of our coordinates such that only the $x$ and $y$
components survive. We consider first the potential (\ref{vec1}).
Giving it a Lorentz boost of magnitude $\beta$ along the positive
z- axis yields
\begin{equation}
{}^{\beta}{A}^I_i={-g \epsilon_{ij} r^j_{\perp} \over r^2_{\perp} }\left [1 -
{{z
- \beta t} \over R_{\beta}} \right ] ~, ~ \label{veb}
\end{equation}
where, $r_{\perp}^2=x^2+y^2$, $R_{\beta} = [(z - \beta t)^2 + (1-
\beta ^2) r_{\perp}^2]^{1\over 2} $ and $i,j=1,2$.
Using the limit \cite{jac}
\[ \lim_{\beta \rightarrow 1}{1\over R_{\beta}} ={1 \over {|x^{-}|}} -
\delta (x^{-})ln \mu ^2 r_{\perp}^2   \label{lim} \]
(where $\mu$ is an arbitrary dimensional parameter, and $x^{\pm}$ are
the usual light cone coordinates), we get
\begin{equation}
\vec{A}~^I_0~\equiv \lim_{\beta \rightarrow 1}~ ^{\beta}A_i~^I = {2g
\over r_{\perp}}~\theta(x^-) {\hat \phi}  \label{bvp}
\end{equation}
which is the vector potential of an ultra-relativistic
monopole. Note that, due to the presence of the
$\theta$ function, $\vec {A}~^I_0$ is singular only along that part of
the z- axis which is {\it behind} the monopole. Similarly, boosting
$\vec{A}~^{II}_0$ one obtains an expression (as $\beta \rightarrow 1$)
which is proportional to $\theta(-x^-)$ and is hence singular along
the z-axis {\it ahead of} the
monopole. In other words, these boosted potentials continue to satisfy
the Wu-Yang criteria \cite{wuy} for non-singular potentials due to a monopole.

The electric and magnetic fields can be calculated from the above to yield
\begin{eqnarray}
B^i ~&=&~  {2g r_{\perp}^i \over {r_{\perp}^2}}{\delta (x^{-})},~~
B^z=0  \nonumber \\
E^i ~&=&~ {2g \epsilon_{ij} r_{\perp}^j \over r_{\perp}^2}{\delta
(x^{-})} ,~~ E^z=0~. \label{fld}
\end{eqnarray}
This shows that $\vec{A}~^I_0$ is a pure gauge everywhere except
on the null plane. Note that the same fields can be
obtained using duality symmetry, i.e. by the
transformation $\vec E \rightarrow \vec B$ and $\vec B \rightarrow
-\vec E$ on the fields already obtained by Jackiw et
al\cite{jac} for the scattering of two point charges. Furthermore, It
is possible to choose a gauge in which only the light cone components
$A_{\pm}$ of the boosted vector potential survive, with the transverse
components vanishing everywhere, yielding the same field strengths as
in eq. (\ref{fld}). We shall discuss this case later.

Next we proceed to compute the scattering amplitude; to this end, we
first rewrite $\vec {A}~^I_0$ as a total derivative in the following form:
\begin{equation}
\vec {A}~^I_0~ =~ 2g \theta (x^{-}) \nabla \phi ~, \label{dif}
\end{equation}
where, $\nabla$ is the gradient operator in the transverse (i.e., $x-y$)
plane. Before the arrival of the monopole with its shock front, the
electric charge is described by a wave function which is a plane wave,
\begin{equation}
\psi _<~ =~\psi _0 ~~~~for~ x^-<0~. \label{bef}
\end{equation}
Immediately after the shock front passes by, the  wave function is
also a plane wave but with a gauge potential dependent phase factor, i.e.,
\begin{equation}
\psi_>~=~exp\left(ie \int dx^{\mu} A_{\mu} \right )~ \psi'_0
\label{aft}
\end{equation}
Upon using eq. (\ref{dif}) for the gauge potential in (\ref{aft}), we obtain
\begin{equation}
\psi _>~ =~ exp[i2eg \phi]~\psi '_0  ~~~~for~ x^->0
\end{equation}
The plane wave solutions $\psi_0$ and $\psi'_0$ are related through
the continuity requirement
\begin{equation}
\psi _< ~=~ \psi _> ~~~~ at~ x^- = 0 .
\end{equation}
Assume now that the monopole-charge system obeys the Dirac quantization
condition
\begin{equation}
eg~~ =~~{ n \over 2}
\end{equation}
so that,
\begin{equation}
\psi _>~ =~ e^{in\phi}\psi '_0~~. \label{pha}
\end{equation}
This sort of phase factor in the almost-forward scattering of a
monopole and a charge was first found by Goldhaber \cite{gol}.
Expanding $\psi$ in plane waves a l\'a 't Hooft \cite{thf}, we get
\begin{equation}
\psi _> = \int dk_+d^2k_{\bot}~A(k_+ , {\vec k}_{\bot} )exp[i{\vec
k}_{\bot} \cdot {\vec r}_\bot -ik_+x_- - ik_-x_+ ] \label{expn}
\end{equation}
with the on shell condition $k_+ = {(k^2_{\bot} + m^2)/k_-}~.$
Multiplying both sides of (\ref{expn}) by a plane wave and integrating
over $x_-$, we have
\begin{equation}
A(k_+,k_\bot) = {\delta (k_+ - p_+) \over (2 \pi)^2}{\int
d^2r_{\bot} exp\left(in\phi + {\vec q} \cdot {\vec r}_\bot \right)}~,
\label{amp}
\end{equation}
where ${\vec q} \equiv {\vec p}_{\bot} - {\vec k}_{\bot}$ is the
transverse momentum transfer, $k$ and $p$ being the final and
initial momenta respectively. After integrating over the angular
variable $\phi$, and appropriate scaling of the radial coordinate,
the integral over the latter in (\ref{amp}) reduces to
$${1 \over q^2} \int_{0}^{\infty} d\rho ~ \rho J_n(\rho)~,$$
where $J_n(\rho)$ is the Bessel function of order $n$. The
integration over the radial coordinate is standard \cite{luke},
yielding
\begin{equation}
\left ({1\over -t}\right )  {2\Gamma (1+ {n \over 2}) \over \Gamma ({n
\over 2})} ~,
\label{rad}
\end{equation}
with $t \equiv -q^2$ being the momentum transfer. Putting everything
together and incorporating canonical kinematical factors, the
scattering amplitude is given by
\begin{equation}
f(s,t)~=~{k_+ \over 2\pi k_0}~{\delta (k_+ - p_+) \left ({n \over
-t}\right)}~. \label{scam}
\end{equation}
Thus, as already mentioned, the scattering amplitude is proportional
to the monopole strength $n$. Note that we would obtain the same
result if we had used the second of the gauge potentials in
(\ref{vec2}) and performed the Lorentz boost etc. One reason for this
is that, the potentials, boosted to $\beta \approx 1$, are both gauge
equivalent to a gauge potential $A'_{\mu}$ given by
$$ {\vec A}'_{\perp}~=~0~=~A'_+~;~A'_-~=~2g \phi
\delta(x^-)~everywhere ~.$$
One can show that this gauge potential produces the same scattering
amplitude as in (\ref{scam}). Thus,
it appears that one can ignore the Dirac string singularity
associated with the monopole gauge potential, in the
kinematical limit under consideration. Observe that this is not true
for boost velocities that are subluminal.

Two remarks are in order at this point. First of all, the scattering
of an ultra-relativistic {\it electric} charge from a
slow-moving Dirac monopole can be shown to yield an identical result
as (\ref{scam}). The easiest way to see this is to use the {\it dual}
formalism wherein one introduces a gauge potential $A_{\mu}^M$ such
that the dual field strength ${\tilde F_{\mu \nu}} \equiv
\partial_{[\mu} A_{\nu]}^M$. If this gauge potential is used to define
electric and magnetic fields, then it follows that the standard field
tensor must satisfy a Bianchi identity, which would then imply that
the gauge potential due to a point charge must have a Dirac
string singularity. Further, the monopole will behave identically to
the point charge of the usual formalism, so that our method above is
readily adapted to produce identical consequences. Secondly, one can
treat the scattering of two Dirac monopoles in the same
kinematical limit exactly as in \cite{jac}, using this dual formalism.
This would yield a
result identical to the one for the electric charge case,
with $e$ and $e'$ being replaced by $g$ and $g'$, the monopole charges.

Next, we consider the effect of turning on the gravitational
interaction between the particles. As shown by 't Hooft
\cite{thf}, the gravitational shock wave due to the fast-moving
particle also produces an extra phase factor in the wave function of
the slow-moving target particle. The net effect of the two shock waves
then is to produce a phase factor that is given by the superposition
of the individual phase factors. It follows that the
integral in eq. (\ref{amp}) must be replaced by
$$\int d^2 r_{\bot}~exp[i(n\phi - Gs~
ln{ \mu ^2 r_{\bot}^2}  +\vec q . \vec r_{\bot})]$$
Once again, the integration over $\phi$ is readily done, giving the
integral
$$ {1 \over q^2} \int_0^{\infty} d \rho ~\rho^{1-2iGs} J_n(\rho)~;~$$
The radial integration is again quite standard, and yields\cite{luke}
an amplitude
\begin {equation}
f(s,t)= {k_+ \over 4\pi k_0}~{\delta(k_+ - p_+)}~({n\over 2}-iGs){\Gamma
({n \over 2}-iGs) \over \Gamma ({n \over 2} + iGs)}\left (4 \over -t \right
)^{1-iGs}~~. \label{gram}
\end{equation}
With this, one can easily show that the cross section
\begin{equation}
{d^2\sigma \over d\vec k_{\perp}^2}~ \sim~ {1 \over t^2} \left ( {n^2 \over4}
+ G^2s^2 \right )~~. \label{csec}
\end{equation}
Clearly, for cm energies of the order of the Planck mass, the
contribution to this cross section due to the monopole-charge
electromagnetic interaction is of the same order of magnitude as that
due to the gravitational interaction between the particles. This is
dramatically different from the case when the particles only have
electric charge, where the electromagnetic
effects constitute a small perturbation on the gravitational effects
at Planck scale.

It is also remarkable that the amplitude given in eq. (\ref{scam})
exhibits poles at $Gs~=~-i(N~+~{n\over 2})$ as opposed to the poles at
$Gs~=~ -iN$ found in the absence of magnetic charge. Following the
argument given in ref.\cite{thf2}, therefore, the spectrum of `bound
states' would appear to admit states of half-odd integral spin (for
odd values of the monopole quantum $n$), in addition to those of
integral spin. This appears to be similar to the spectrum discerned
in three-particle scattering amplitudes \cite{thf2}. While it
is well known \cite{saha} that a monopole-charge  system
obeying Dirac quantization carries half-odd
integral spin, the relation of this to the spectrum of states above is
not quite transparent. In any event, the existence of such states in
the spectrum would tend to imply that the underlying string structure
(if any) corresponds to a {\it supersymmetric} rather than a
bosonic string. Perhaps this would become clearer if we were to follow
Verlinde and Verlinde \cite{ver} in an attempt to identify a topological sigma
model for the pure gauge states of the system. We hope to report on
this elsewhere.

\end{document}